\documentclass[twocolumn,english,aps,prl,floatfix,showpacs]{revtex4}
\usepackage[T1]{fontenc}
\usepackage[latin1]{inputenc}
\usepackage{graphicx}
\usepackage{amssymb}

\makeatletter

\usepackage{graphicx}
\usepackage{amssymb}

\usepackage{babel}

\usepackage{babel}
\makeatother
\begin{document}

\title{Superconducting States of Pure and Doped Graphene}

\author{Bruno Uchoa, and A. H. Castro Neto}

\affiliation{Physics Department, Boston University, 590 Commonwealth Ave., Boston,
MA 02215}

\begin{abstract}
We study the superconducting phases of the two-dimensional honeycomb lattice 
of graphene. We find two spin singlet pairing states, $s$-wave and 
an exotic $p+ip$ that is possible because of the special structure of the
honeycomb lattice. At half filling, the $p+ip$ phase is gapless and superconductivity
is a hidden order. We discuss the possibility of a superconducting state
in metal coated graphene. 
\end{abstract}


\maketitle

Graphene is a two-dimensional (2D) electronic system on a honeycomb lattice
whose electronic excitations can be described in terms of linearly dispersing 
Dirac fermions \cite{Novoselov}. Because of its unusual properties \cite{Peres}, such as
anomalous integer quantum Hall effect \cite{Novoselov2,Zhang} and 
universal conductivity \cite{Novoselov2}, graphene has attracted a
lot of attention in the condensed matter community. One of the
interesting properties of graphene is that its chemical potential
can be tuned through a electric field effect, and hence it is possible
to change the type of carriers (electrons or holes), opening the
doors for a carbon based electronics. Superconductivity 
has been induced in short graphene samples through proximity effect with 
superconducting contacts \cite{proxEffect}. This indicates that Cooper pairs 
can propagate coherently in graphene.  From the theoretical side, 
anomalous Andreev reflection \cite{beenakker} and transport 
\cite{titov} have been predicted in graphene
junctions with superconductors. It is known that the electronic properties of
graphene are modified by changing the number of graphene planes
\cite{stacks}.
In particular, bilayer graphene has been demonstrated to be a tunable
gap semiconductor \cite{castro}. These results raise the question
of whether it would be possible to modify graphene, either structurally or
chemically, so that it would become a magnet \cite{ferro} or even an 
intrinsic superconductor. 
By exploring the number of graphene layers
and the chemical composition, it maybe possible to tailor its electronic
properties.    

\begin{figure}
\begin{centering}\includegraphics[scale=0.3]{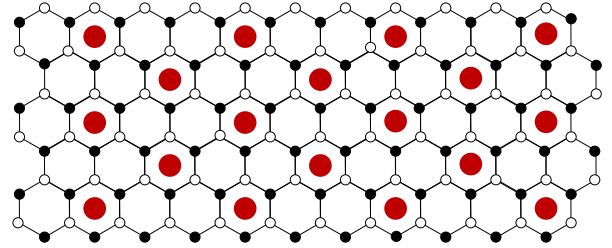}\par\end{centering}
\caption{\label{CaC6}{\small (color online) Graphene coated with metal. 
Small black (white) circles belong to the A (B) sublattices;
Large (red) circles represent the metal atoms. }}
\end{figure}

In this letter, we derive a mean-field phase diagram for spin singlet superconductivity in 
graphene. We show that besides the usual s-wave pairing, 
an unexpected spin singlet state with $p+ip$ orbital symmetry is possible
because of the structure of the honeycomb lattice. In fact, we  
show that the $p+ip$ state is preferred to the s-wave state if the on-site
electron-electron interactions are repulsive. $p+ip$-pairing
states are rather interesting \cite{annett} because, in the presence of a magnetic field, the
superconducting vortices are described in terms of Majorana fermions 
with non-abelian statistics that can be used
for topological quantum computing \cite{ivanov}. 
Although it is very hard to predict
pairing mechanisms from microscopic models, we examine possible phonon and
plasmon mediated superconductivity in chemically modified, metal coated, 
graphene, as shown in Fig.~\ref{CaC6}. Our results indicate that a plasmon
mediated superconductivity is possible in this system.

Notice that on symmetry grounds an electronic state on a honeycomb lattice has three
different components associated with the continuous SU(2) spin symmetry and the
discrete lattice symmetry (the honeycomb lattice can be described as a
triangular lattice with a basis with two atoms, A and B, as in
Fig.~\ref{CaC6}). At low energies and long wavelengths, close to the Dirac point, the system has
effective SO(3) $\otimes$ Z$_2$ spatial symmetry. 
Hence, the superconducting state can have a Cooper pair wavefunction of the form
$
\Psi_{{\rm pair}} = \psi_{S} \otimes \psi_{L} \otimes \psi_{A-B} \, ,
$
where $\psi_S$ is the spin, $\psi_L$ is the orbital, and
$\psi_{A-B}$ is the sub-lattice component. Pauli's principle requires the
wavefunction to be anti-symmetric for the exchange of particles. 
For a spin singlet state, $S=0$, one can have either $L= {\rm even}$ and $A-B$
symmetric ($L=0$ being the s-wave), or $L={\rm odd}$ and $A-B$ anti-symmetric
($L=1$ being p-wave). 

The free electron Hamiltonian can be written as:
\begin{equation}
H_{t}=-\mu\sum_{i}\hat{n}_{g,i}-
t\sum_{\langle
  ij\rangle}\sum_{s=\uparrow\downarrow}(a_{i,s}^{\dagger}b_{j,s}+h.c.)\,,
\label{Ht}\end{equation}
 where $t\approx2.8$ eV is the hopping energy between nearest neighbor
C atoms, $a_{i,s}$ ($a_{i,s}^{\dagger}$) is the on-site annihilation
(creation) operator for electrons in the sublattice $A$ with spin
$s=\uparrow,\downarrow$, and $b_{i,s}$ ($b_{i,s}^{\dagger}$) for
sublattice $B$, $\hat{n}_{g,i}$ is the on-site particle density
operator, and $\mu$ is the graphene chemical potential (we use units such that $\hbar=1=k_{B}$).
Diagonalization of (\ref{Ht}) leads to an spectrum given by: 
$\varepsilon_{\mathbf{k}}=-t|\gamma_{\mathbf{k}}|$,
where $\mathbf{k}$ is the 2D momentum, and 
$\gamma_{\mathbf{k}}=\sum_{\vec{\delta}}\textrm{e}^{i\mathbf{k}\cdot\vec{\delta}}$
($\vec{\delta}_{1}=a(\hat{x}/2+\sqrt{3}/{2}\hat{y})$, $\vec{\delta}_{2}=a(\hat{x}/2-\sqrt{3}/{2}\hat{y})$,
and $\vec{\delta_{3}}=-a\hat{x}$, where $a\approx1.42$ \AA \,
is the C-C distance). At the corners of the hexagonal Brillouin
zone (at $\mathbf{Q}_{0}=[0,\pm 4\pi/(3\sqrt{3}a)]$), 
the band has the shape of a Dirac cone: 
$\varepsilon_{\mathbf{Q}_{0}+\mathbf{k}}=\pm v_{0}|\mathbf{k}|$,
where $v_{0}=3at/2\approx6$ eV \AA \, is the Fermi-Dirac velocity.
In neutral graphene, the chemical potential crosses exactly through
the Dirac point ($\mu=0$). 

\begin{figure}
\begin{centering}\vspace{-0.4cm}\par\end{centering}
\begin{centering}\includegraphics[scale=0.18]{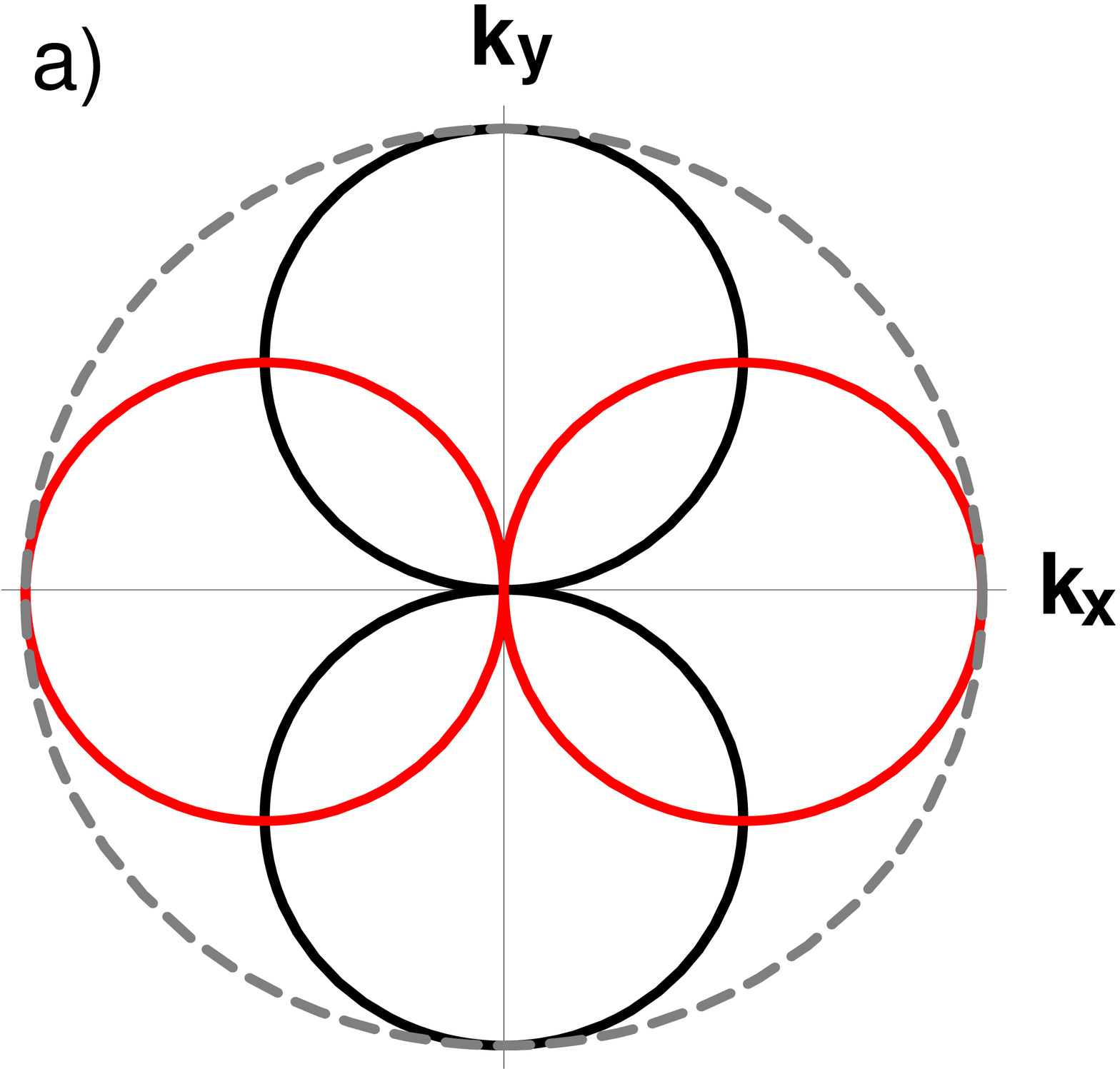}$\qquad$\includegraphics[scale=0.18]{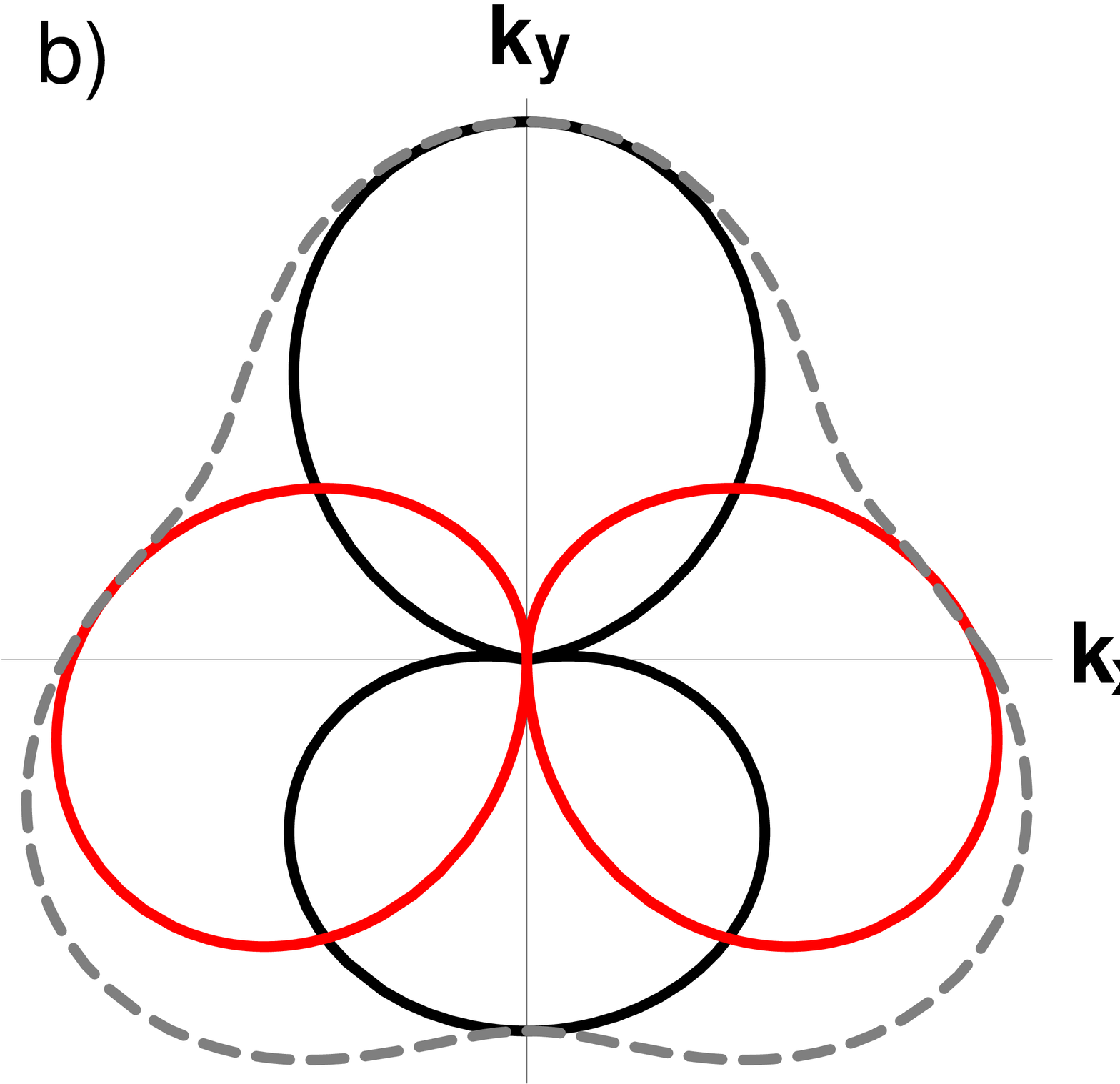}
\par\end{centering}
\caption{\label{pip}{\small (color online) $p+i p$ order parameter in momentum space: 
(a) close to the Dirac point and (b) away from it. Dark (black) line is
the real part, gray (red) is the imaginary part, and dashed is 
the amplitude of the order parameter.}}
\end{figure}

The electron-electron interactions are described by:
\begin{eqnarray}
H_{P} & = & \frac{g_{0}}{2}\sum_{is}\left[a_{is}^{\dagger}a_{is}a_{i-s}^{\dagger}a_{i-s}+b_{is}^{\dagger}b_{is}b_{i-s}^{\dagger}b_{i-s}\right]\nonumber \\
 &  & +g_{1}\sum_{\langle ij\rangle}\sum_{s,s^{\prime}}a_{is}^{\dagger}a_{is}b_{js^{\prime}}^{\dagger}b_{js^{\prime}}\,,
\label{Hefij}
\end{eqnarray}
where $g_{0}$ and $g_{1}$ are on-site and nearest neighbor
electron-electron interaction energies, respectively. 
It is easy to see that the  superconducting order parameters for spin singlet
 are: (1) s-wave: $\Delta_{0}=\langle a_{i\downarrow}a_{i\uparrow}\rangle=\langle b_{i\downarrow}b_{i\uparrow}\rangle;$
(2) $p$-wave: 
$\Delta_{1,ij}=\langle a_{i\downarrow}b_{j\uparrow}-a_{i\uparrow}b_{j\downarrow}\rangle\,.$
We assume $\Delta_{1,ij}=\Delta_{1}$ for all nearest neighbors and zero,
 otherwise. In the momentum space one has: 
$
\Delta_{\mathbf{k}}=\sum_{ij}\Delta_{1,ij}\textrm{e}^{-i\mathbf{k}\cdot(\mathbf{r}_{i}-\mathbf{r}_{j})}\:=\:\Delta_{1}\gamma_{\mathbf{k}}^{*}\,.
$
 Close to the Dirac points $\mathbf{Q}_{0}$ (Fig.~\ref{pip}(a)) the order
parameter can be written as, 
 $\Delta_{\mathbf{Q}_{0}+\mathbf{k}}=(3a/2)\Delta_{1}(k_{y}+ik_{x})$,
that is, it has $p+i p$ symmetry. At high energies, away from the Dirac
point, the discrete symmetry of the lattice is recovered and the pairing state
is modified as in Fig.~\ref{pip} (b). For simplicity, however, we refer to this
phase as $p+ip$, which is the symmetry close the Dirac cone. 

Decoupling the interaction terms in (\ref{Hefij}) gives,  
\begin{eqnarray}
H_{P} & = &
E_{0}+g_{0}\Delta_{0}\sum_{i}\left(a_{i\uparrow}^{\dagger}a_{i\downarrow}^{\dagger}+b_{i\uparrow}^{\dagger}b_{i\downarrow}^{\dagger}\right)+{\textrm{h.c.}}
\nonumber 
\\
 &  & +g_{1}\sum_{\langle
   ij\rangle}\Delta_{1,ij}\left(a_{i\uparrow}^{\dagger}b_{j\downarrow}^{\dagger}-a_{i\downarrow}^{\dagger}b_{j\uparrow}^{\dagger}\right)+{\textrm{h.c.}} \, ,
\label{HP}
\end{eqnarray}
and the total Hamiltonian can be diagonalized via a Bogoliubov transformation: 
$H_{{\rm eff}}=\sum_{\mathbf{k},\alpha,s} \omega_{\mathbf{k} \alpha s}\hat{n}_{\mathbf{k}\alpha s}^{B}+E_{0}\,,$
where $E_{0}=-g_{0}\Delta_{0}^{2}-3g_{1}\Delta_{1}^{2}$ ($\Delta_0$ and
$\Delta_1$ are real numbers), and $\hat{n}_{\mathbf{k},\alpha}^{B}$
is the quasi-particle number operator. The spectrum is: 
$\omega_{\mathbf{k},\alpha,s}\equiv \alpha \, \omega_{\mathbf{k},s}$, with $\alpha,s = \pm 1$
and, 
\begin{eqnarray}
\omega_{\mathbf{k},s} & = &
\sqrt{\left(t|\gamma_{\mathbf{k}}|+s\mu\right)^{2}+\left(g_{0}\Delta_{0}+sg_{1}\Delta_{1}
|\gamma_{\mathbf{k}}|\right)^{2}} \, .
\label{disp}
\end{eqnarray}
For $\Delta_{0}\neq0$
and $\Delta_{1}=0$, eq. (\ref{disp}) describes an s-wave state with a 
gap given by: $E_g^{\{0\}}=2|g_{0}\Delta_{0}|$. In the  case,
$\Delta_{0}=0$ and $\Delta_{1}\neq0$, the system has an isotropic gap 
$
E_g^{\{1\}} =  2|\mu g_{1}\Delta_{1}|/\sqrt{t^{2}+g_{1}^{2}\Delta_{1}^{2}}\,,
$
which scales linearly with $\mu$. In the neutral limit ($\mu=0$) for
$\Delta_{0}=0$, 
the dispersion (\ref{disp}) is gapless, with
$\omega_{\mathbf{k},s}=\bar{t}|\gamma_{\mathbf{k}}|$, 
where $\bar{t}$ is the effective hopping energy $\bar{t}=t\sqrt{1+g_{1}^{2}\Delta_{1}^{2}/t^{2}},$
which renormalizes the Fermi-Dirac velocity. Notice that in this case, the
superconducting state does not lead to a gap in the spectrum but to a
renormalization of the velocity. This state of affairs we call hidden order.
In the case for $\Delta_{0}\neq0$
and $\Delta_{1}\neq0$, particle-hole symmetry is broken and the gap 
is given by $E_g^{\{0,1\}}=2|tg_{0}\Delta_{0}- g_{1}\mu\Delta_{1}|/\bar{t}$. 

\begin{figure}
\begin{centering}\includegraphics[scale=0.25]{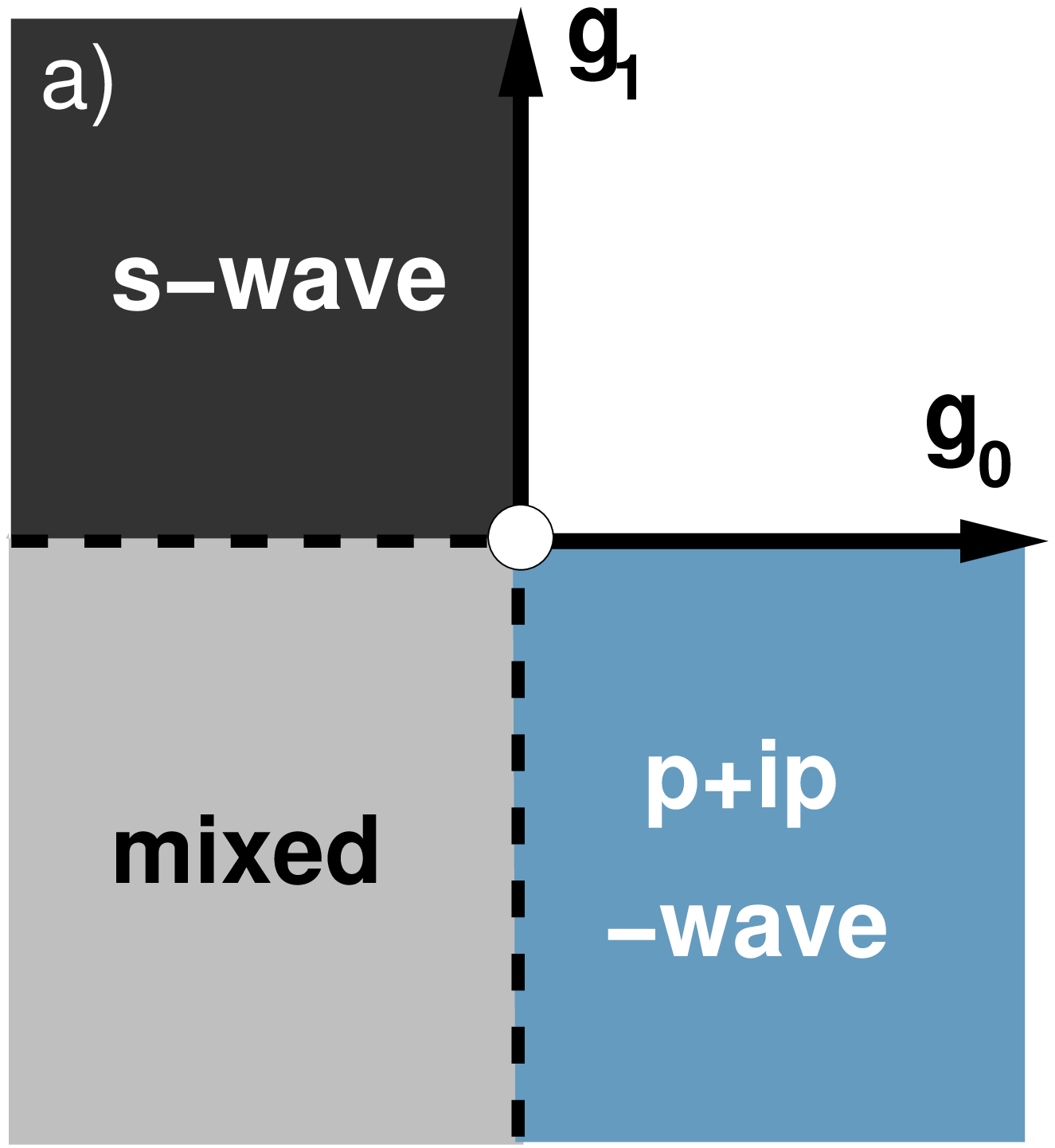}$\qquad$\includegraphics[scale=0.25]{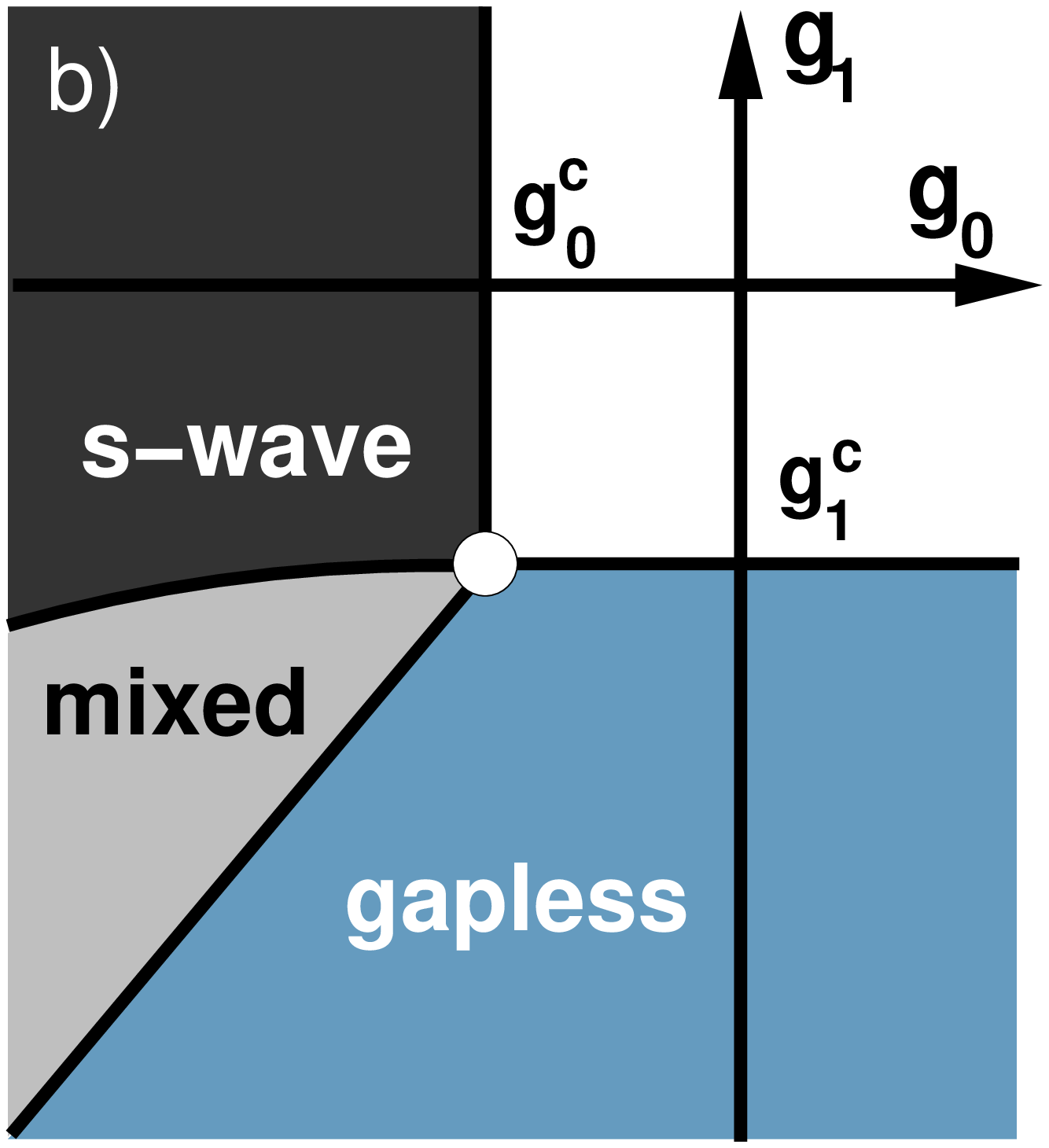}\par
\end{centering}
\caption{\label{phasediag}{\small Mean-field phase diagram: 
(a) $\mu \neq 0$; (b) and $\mu=0$.
Dashed lines are first order transitions, continuous lines
are second order.}}
\end{figure}

The values of $\Delta_{0}$ and $\Delta_{1}$ are calculated by minimizing 
the free energy: $F=-\frac{1}{\beta}
\sum_{\mathbf{k},\alpha,s}\ln\left(1+\textrm{e}^{-\beta\omega_{\mathbf{k}\alpha
    s}}\right)+E_{0},$
where $\beta=1/T$. The coupled self-consistent equations for the
order parameters are: 
\begin{eqnarray}
\Delta_{0} \!\!\!& = &\!\!
-\sum_{\mathbf{k},s} \! \left(g_{0}\Delta_{0}\!+\!s|\gamma_{\mathbf{k}}|g_{1}\Delta_{1}\right)\!
\tanh\!\left(\beta\omega_{\mathbf{k}s}/2\right)/(2\omega_{\mathbf{k}s})
\, ,
\nonumber
\\
\Delta_{1} \!\!& = &\!\!\!
-\sum_{\mathbf{k},s} |\gamma_{\mathbf{k}}|\left(g_{1}\Delta_{1}|\gamma_{\mathbf{k}}|\!+\!sg_{0}\Delta_{0}\right)
\! \tanh\!\left(\beta\omega_{\mathbf{k}s}/2\right)/(6\omega_{\mathbf{k}s}) 
\nonumber
\, .
\end{eqnarray}
For $\mu \neq 0$, one finds
three distinct phases (see Fig.~\ref{phasediag} (a)): 
(\emph{i}) a $s$-wave phase for attractive on-site ($g_{0}<0$) and repulsive
nearest neighbor ($g_{1}>0$) interactions; (\emph{ii}) a $p+ip$ phase
for repulsive on-site ($g_{0}>0$) and attractive nearest neighbor
interactions ($g_{1}<0$); and (\emph{iii}) a co-existence phase for fully
attractive interactions ($g_{0},g_{1}<0$). The superconducting transitions from
normal to $s$-wave, and normal to $p+ip$ are of second order. 
The transitions involving the mixed phase for $g_{0}<0$ and $g_{1}\rightarrow0$
or $g_{1}<0$ and $g_{0}\rightarrow0$ are of first-order even at
$T=0$. At the critical temperature, $T_{c}$, the phase transitions
of phases (\emph{i}) and (\emph{ii}) to the normal state are second order,
while the transition between the mixed phase and the normal phase
is abrupt. In the weak coupling limit of phase (\emph{i}), i.e. for  $|g_{0}|\ll- g_0^c\equiv \pi v_{0}^{2}/\Lambda$,
where $\Lambda$ is a high energy cut-off, the critical
temperature is given by \cite{Uchoa}:
$
T_{c}\approx2\mu(\gamma/\pi)\exp\{-\Lambda(g_{0}^{c}/g_{0}-1)\mu^{-1}-1\},
$
where $\ln\gamma\sim0.577$ is the Euler constant. In reality, however,
because the system is 2D there can be no true superconducting long-range
order (Mermin-Wagner theorem) but there will be a Kosterlitz-Thouless
(KT) transition below a certain temperature $T_{KT}<T_{c}$. Hence,
$T_{c}$ only establishes the temperature below which the amplitude
of the order parameter becomes finite while its phase still fluctuates.
Phase coherence only occurs at low temperatures and depends on the
phase stiffness of the system. Although the transition is of the KT
type, one expects a precipitous drop of the resistivity of the material
below $T_{KT}$ indicating the entrance of the electrons into a state
of quasi-long-range superconducting order.

For $\mu=0$ superconductivity
requires a minimum coupling to occur (the problem becomes quantum critical). 
The quantum critical lines are given by
$g_{0}=g_{0}^{c}\equiv-\pi v_{0}^{2}/\Lambda$ and $g_{1}=g_{1}^{c}\equiv-4\pi v_{0}^{4}/(a^2\Lambda^{3})$,
as shown in Fig. \ref{phasediag} (b). We identify the phases: (\emph{iv})
$s$-wave for $g_{0}<g_{0}^{c}$ and $g_{1}>\, f(g_{0})\equiv g_{0}^2 g_{1}^{c}/[g_{0}^2-(3/2)(g_{0}-g_{0}^{c})^{2}]$;
(\emph{v}) gapless phase, for
$g_{1}<h(g_{0})\equiv4[v_{F}^{2}/(a^2\Lambda^{2})]g_{0}<g_{1}^{c}$; 
and (\emph{vi}) mixed phase, for $h(g_{0})<g_{1}<f(g_{0})$,
where the symmetry is mixed between $s$ and $p+ip$ pairing.

We remark that the physical realization of superconductivity
in neutral graphene is difficult because of the vanishing  density of 
states and the absence of electron-electron screening. Therefore, in order
to superconductivity to develop easily one has to substantially 
shift the graphene
chemical potential away from the Dirac point. This can be achieved
by chemically doping graphene with a metal coating: when an alkaline metal 
is placed on top of a graphene crystal, the $s$
electrons migrate to the $\pi$-band to compensate the strong difference
in electro-negativities, raising up the chemical potential from the
Dirac points and lowering the energy of the metallic bands in order
to establish electrostatic equilibrium. Since neither carbon nor
alkaline metals alone superconduct in ordinary conditions, 
we identify two main possible
mechanisms for superconductivity on coated graphene: 1) \emph{electron-phonon}
coupling of the ions with the electrons in the modified metallic bands,
and 2) \emph{electron-plasmon} coupling of the graphene electrons
with the acoustic plasmons of the metal. 

The electron-phonon mechanism tends
to favor superconductivity at {\it high} electronic densities, and
has been used to explain the bulk superconductivity of graphite intercalated
CaC$_{6}$ \cite{Lamura,Mazin,Calandra}. The strength of
the electron-phonon coupling can be extracted from
\emph{ab initio} calculations and from experimental data, which are
not currently available for coated graphene. Nevertheless, this mechanism
is conventional and we will not discuss it here.

In the electron-plasmon mechanism, the
attractive electron-electron interaction is mediated by a screened acoustic plasmon
of the metal, which is favorable to superconductivity at {\it low} 
electronic densities, where the plasmon
is weakly damped by the graphene particle-hole continuum. In the
remaining of the letter we investigate under what conditions this
mechanism could be effective for graphene superconductivity. Since
electrons and acoustic plasmons have comparable energy scales, a reliable
calculation of the critical temperature requires the study of retardation
effects that we will cover in future publications.

Plasmon mediated superconductivity has been widely studied in the
past \cite{Frohlich,Richarson,Bill,Ruvalds,Kresin} and we concentrate
on the particular aspects of the graphene problem.
In coated graphene, the Coulomb interaction between the layers induces
an effective electron-electron interaction for electrons in the graphene
layer that can be calculated with the use of the random phase approximation
(RPA) \cite{Pines2}. The RPA expansion in terms of the zeroth order polarization functions
of the metal, $\Pi_{m}^{0}$, and of graphene, $\Pi_{g}^{0}$, results
in an effective \textit{retarded} interaction of the form: 
\begin{eqnarray}
H_{ef}^{g}=\sum_{\mathbf{q}}\sum_{\omega}V_{ef}(\mathbf{q},\omega)\,\hat{n}_{g}(\mathbf{q},\omega)\hat{n}_{g}(-\mathbf{q},-\omega)\,,\quad
\label{Hef}
\end{eqnarray}
where, 
\begin{eqnarray}
V_{ef}(\mathbf{q},\omega) \!=\!
V_{0,q}/\epsilon(\mathbf{q},\omega)\left[1\!-\!(V_{0,\mathbf{q}}\!-\!V_{d,\mathbf{q}})\Pi_{m}^{0}(\mathbf{q},\omega)\right]\!,\quad
\label{Veftotal}
\end{eqnarray}
is the effective electron-electron interaction, and
\begin{eqnarray}
\epsilon(\mathbf{k},\omega) & = &
1-V_{0,q}\left[\Pi_{g}^{0}(\mathbf{k},\omega)+\Pi_{m}^{0}(\mathbf{k},\omega)\right]
\nonumber 
\\
 &  &
 +(V_{0,q}^{2}-V_{d,q}^{2})\Pi_{m}^{0}(\mathbf{k},\omega)\Pi_{g}^{0}(\mathbf{k},\omega)\qquad\qquad
\label{epsilonT}
\end{eqnarray}
 is the total dielectric function of the system, with 
$V_{d,\mathbf{q}}=2\pi e^{2}\textrm{e}^{-qd}/(\epsilon_{0}q)$
as the Fourier transform of the Coulomb interaction between electrons
in two layers separated by a distance $d$. In Eq. (\ref{epsilonT}),
the separation between the metal and graphene layers induces a cross
polarization term between the two layers that vanishes when $d=0$.
In the opposite limit, $kd\gg1$, the two layers decouple. 

The electronic susceptibility of metals is commonly described
by the Lindhard polarization function. When
$\omega>v_{F}q$, where $v_{F}$ is the Fermi velocity of the metal,
the $q\rightarrow0$ limit of the 2D Lindhard function gives \cite{Pines2}
$V_{0,q}\Pi_{m}^{0}(q,\omega)=\Omega_{m}^{2}/\omega^{2}\,,$ where
$\Omega_{m}(q)=e\sqrt{2E_{F}q/\epsilon_{0}}$ is the plasmon
of the 2D electron gas,
and $E_{F}$ is the Fermi energy of the metallic band. The polarization
function of graphene at small momentum $q$ is dominated by intra-band
excitations connecting states in the same branch of the Dirac cone.
At lowest order, it has the same momentum and frequency dependence
of the polarization function of an infinite stack of graphite layers
in the absence of interplane hopping \cite{Shung}: 
\begin{eqnarray}
\Pi_{g}^{0}(q,\omega) & = & -(2\mu/\pi
v_{0}^{2})\left[1-\omega/\sqrt{\omega^{2}-v_{0}^{2}q^{2}}\right].
\label{Pig}\end{eqnarray}

\begin{figure}[b]
\begin{centering}\includegraphics[scale=0.25]{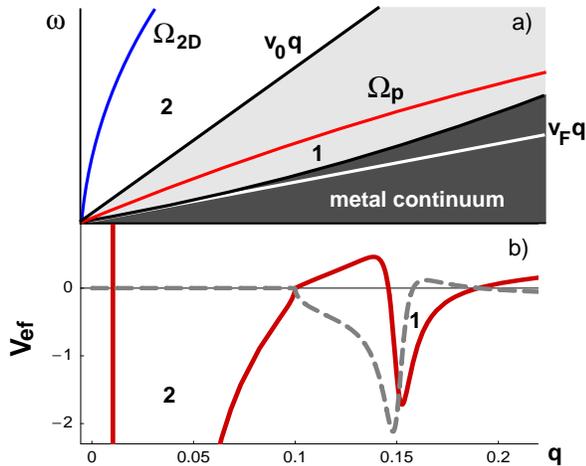} \par
\end{centering}
\caption{\label{omegaq}{\small a) Low energy excitations in the metal-graphene
system. Dark and grey regions are the particle-hole continuum of the
metal and of graphene, respectively. b) Real (solid) and imaginary
(dashed) parts of the effective interaction (\ref{Veftotal}), in
units of $2\pi e^2 v_{0}/(\epsilon_{0}\mu)\sim$ 45 eV \AA$^{2}$
vs. momentum normalized by $\mu/v_{0}$, for $\omega=0.1\mu$, $\mu=2$
eV, $E_{F}=0.4$ eV.}}
\end{figure}

In the low frequency limit, $\omega\ll v_{0}q$, the polarization
function of graphene is approximated by its static part, $\Pi_{g}^{0}(q,0)$,
and the total dielectric function (\ref{epsilonT}) becomes 
$\epsilon(q,\omega)=\epsilon_{g}(q,0)\left[1-\Omega_{p}^{2}(q)/\omega^{2}\right]\,,$ where, 
\begin{equation}
\Omega_{p}^{2}(q)\!\approx\! \Omega_{m}^{2}(q)/\epsilon_{g}(q,0) \left[1\!-\!\left(1\!-\!\textrm{e}^{-2qd}\right)\!V_{0,q}\Pi_{g}^{0}(q,0)\right],
\label{screenedPlasmon}
\end{equation}
is the screened plasmon mode of the metal (i.e., $\epsilon(q,\Omega_p(q))=0$). When $qd\ll1$ and
$|\omega|\ll v_{0}q$, this plasmon has a linear dispersion, $\Omega_{p}(q)\approx\sqrt{E_{F}^{*}/(2\mu)}v_{0}q$,
where $E_{F}^{*}=E_{F}\left[1+(8e^{2}d\mu/\epsilon_{0}v_{0}^{2})\right]$.
For $qd\gg1$, the plasmon is not screened by the graphene layer,
and one recovers the plasmon dispersion for the 2D electron gas: 
$\Omega_{p}\rightarrow\Omega_{m}\propto\sqrt{q}$.
In the region $v_{F}q<|\omega|\ll v_{0}q$, eq.~(\ref{Veftotal})
can be approximated as: 
\begin{eqnarray}
V_{ef}(q,\omega)\!\!\approx\!\! 
V_{0,q}/\epsilon_{g}(q,0)\!\left[\Omega_{p}^{2}/(\omega^{2}-\Omega_{p}^{2})+1\right]
\, ,
\label{Vef}
\end{eqnarray}
In the attractive region 1 of Fig.~\ref{omegaq}, the electrons
of graphene screen the charge fluctuations, restoring the longitudinal
response of the collective modes in the normal phase, that is required
to preserve the local gauge invariance of the superconductor in the London limit \cite{Pines}.
The electrons of the metal by their turn are slow and tend to anti-screen
the graphene electrons \cite{Garland}, producing an average repulsive
interaction in the metallic band. In the high frequency limit $|\omega|\gg v_{0}q$,
the effective interaction (\ref{Veftotal}) has a second region of
attraction (region 2 of Fig.~\ref{omegaq}). For $qd\ll1$, the effective
interaction in this region can be approximated by 
$V_{ef}(q,|\omega|\gg v_{0}q)=V_{0,q}\left[\Omega_{2D}^{2}/(\omega^{2}-\Omega_{2D}^{2})+1\right]$,
where $\Omega_{2D}(q)=e\sqrt{2(E_{F}+\mu)q/\epsilon_{0}}$ is the
plasmon dispersion of the 2D electron gas. Although the interaction is attractive
for $|\omega|<\Omega_{2D}(q)$, this attractive region is present
in the ordinary 2D electron gas and it does not lead itself to superconductivity,
although it favors superconductivity by reducing the Coulomb
repulsion. 

A necessary condition for appearance of plasmon induced superconductivity
is that the screened acoustic plasmon is not overdamped
by the particle-hole continuum in the interval $v_{F}q<\Omega_{p}\lesssim v_{0}q$.
In the long wavelength limit, $qd\ll1$, the condition for the existence
of this acoustic mode is $E_{F}\lesssim2\mu\epsilon_{0}/(\epsilon_{0}+8\alpha_{g}d\mu/v_{0})\;<\; mv_{0}^{2}/2$
$\approx2.4$ eV. For $\mu\sim2$eV, $\epsilon_{0}\sim5$ and $d\sim3$ \AA \,
\cite{Lamoen}, the left hand side of the inequality becomes $E_{F}\lesssim1$
eV, or equivalently that the electronic concentration in the metal
layer has to be smaller than $\sigma_{c}\approx4\times10^{14}$ electrons
cm$^{-2}$ (or $0.12$ electrons per C). If $x$ is the number of
remaining electrons per metallic atom M, for a system with chemical
composition $[{\textrm{M}}_{n}{\textrm{C}}_{m}]_{2D}$, one of the
conditions is that $x\lesssim0.12(m/n)\,.$ For K coated graphene
$[{\textrm{K}}{\textrm{C}}_{8}]_{2D}$, which is known to form a stable
metallic lattice \cite{Caragiu}, the condition is $x\lesssim0.96$
electrons per K. Values of $x$ $\sim0.54$$-0.83$ were obtained
by \emph{ab initio} calculations for K adsorbed in graphite \cite{Caragiu}.

In conclusion, we have derived the mean-field phase diagram
for superconductivity in a honeycomb lattice, where a novel singlet
$p+ip$ phase appears. We have examined the mechanism of plasmon
mediated superconductivity in graphene with the aim of proposing new carbon
low-dimensional systems where superconductivity can be observed. 

We thank V. Falko, 
F.~Guinea, N.~M.~R.~Peres, and S.-W. Tsai for many discussions
and comments. This work was supported by NSF grant DMR-0343790. B.
U. acknowledges CNPq, Brazil, for the support under the grant 201007/2005-3.

\end{document}